# Most Ligand-Based Classification Benchmarks Reward Memorization Rather than Generalization

**Izhar Wallach**[*]
Atomwise, Inc.
221 Main St.
San-Francisco, CA, 94105
izhar@atomwise.com

**Abraham Heifets**
Atomwise, Inc.
221 Main St.
San-Francisco, CA, 94105
abe@atomwise.com

## Abstract

Undetected overfitting can occur when there are significant redundancies between training and validation data. We describe AVE, a new measure of training-validation redundancy for ligand-based classification problems that accounts for the similarity amongst inactive molecules as well as active. We investigated seven widely-used benchmarks for virtual screening and classification, and show that the amount of AVE bias strongly correlates with the performance of ligand-based predictive methods irrespective of the predicted property, chemical fingerprint, similarity measure, or previously-applied unbiasing techniques. Therefore, it may be that the previously-reported performance of most ligand-based methods can be explained by overfitting to benchmarks rather than good prospective accuracy.

## Introduction

The computational chemistry literature paints a rosy picture of achievable predictive performance. Today, it is typical to see predictive accuracies in excess of 0.8 AUC (Area under the Receiver Operator Characteristic curve)[1;2] [1] with some reports even above 0.99 AUC on standard benchmarks[2]. If predictive models truly and consistently generalized at such accuracies, it would also mean that – like physics before us – we would have created models that are more accurate than experiments[3;4]. Unfortunately, accuracy in practice is not as good as benchmark results suggest; gold standard data in biochemistry is still measured rather than computed. Therefore, it is necessary to investigate whether our field's benchmarks contain design flaws that yield overly optimistic performance estimates. Accurate self-assessment is critically important to computational chemistry practitioners, collaborators, and skeptics.

Effective benchmark design is subtle but important. One source of benchmark failure is redundancy between the training and validation sets. In the limit, when the training and validation sets are the same, a system that memorizes the training data achieves perfect benchmark performance while completely failing to generalize to new data. Far from being a success of the predictive system, such performance is correctly interpreted as a failure in benchmark design.

Instead, a benchmark must typify the intended use of the evaluated system. One common use is making predictions for data not seen during training, so benchmark performance should reflect generalization to novel data (with the kind of "novelty" that the system might reasonably be expected to encounter in practice). For example, two of the most widely-used machine learning benchmarks, MNIST[5] for handwriting recognition and TIMIT[6] for speech transcription, both eschew random partitioning of data into training and validation sets. MNIST splits training and validation by author, rather than by example. That is, the benchmark design captures the fact that a system for hand-writing recognition will encounter authors missing from the training set. Similarly, the TIMIT design specifies that the validation set should comprise speakers not in the training data and maintain a balance between male and female voices. In the context of YouTube recommendations, Covington et al.[7] discuss how temporal splitting rather than random partitioning of training data is necessary to prevent overfitting due to correlations in user video watching behavior. In each of these examples, a random partition of available data leaves a validation set with unreasonable similarity to the training data and, therefore, unreasonably optimistic benchmark performance.

---

[*]Corresponding Author

[1]A 0.8 AUC corresponds to an 80% chance that a randomly-chosen active molecule will be ranked above a randomly-chosen inactive.

Chemical data is rife with potential redundancies and biases. Experiments do not sample chemical space uniformly, and the resulting patterns may dominate benchmark measurements without capturing whether the system is providing predictive value for novel investigations. Only a small and biased portion of chemical space has ever been synthesized[8] and, even within discovery projects, medicinal chemistry efforts cluster near the current best molecules[9]. Molecules may be selected or excluded for legitimate reasons that are unrelated to their coverage of chemical space such as cost, synthetic feasibility, availability ("already in the library"), or replication of previous experimental results as positive or negative controls.

Ignoring these factors during benchmark design risks constructing a validation test which fails to identify which systems have useful predictive accuracy. Good et al.[10] show datasets can suffer from *analog bias* such that memorizing over-represented scaffolds leads to un-realistically good benchmark performance. Verdonk et al.[11] previously described *artificial enrichment* and Ripphausen et al.[12] described *complexity bias*, in both cases where active molecules are easily differentiated from inactives by differences in coarse properties such as molecular weight or number of hydrogen bond donors. Kearnes et. al[13] demonstrated that random cross-validation leaves many identical molecules between training and validation sets, which encourages overfitting and skews any eventual performance measurements. Cleves et al.[14] and Jain et al.[15] describe *inductive bias* and *confirmation bias*, respectively, which note that molecules are built by human chemists using human judgment and are often deliberately biased to resemble molecules previously shown to exhibit desirable characteristics.

A number of techniques have been previously introduced to mitigate bias and data redundancy in computational chemistry benchmarks. Unterthiner et al.[1] used single-linkage clustering to group molecules into similar scaffolds, and then partitioned the clusters into different cross-validation folds. Junshui et al.[16] partitioned data temporally, with data generated before a given date comprising the training data and data after that date comprising the validation; however, Kearnes et al.[13] subsequently showed that, even when time-splitting does not leak information across partitions, a significant number of validation molecules can be near-duplicates of training molecules. Rohrer and Baumann[17] proposed a method for selecting Maximally Unbiased Validation sets (MUV) that attempt to ensure that actives are uniformly embedded within inactives with respect to some similarity metric. Xia et al.[18] proposed a similar method where they not only enforce active ligand diversity but also make sure that pairs of actives and pairs of property-matched decoys[19] have the same range of fingerprint similarities.

Despite these previous attempts, we find that a wide range of benchmarks in the field still suffer from redundancy and bias. In this paper, we introduce a new simple redundancy measure and demonstrate that the measure is predictive of machine learning performance across a variety of machine learning algorithms, feature definitions, distance functions, benchmarks, and previously-applied unbiasing techniques.

Specifically, we propose a new approach inspired by the MUV unbiasing technique[17]. MUV unbiasing defines two functions *nearest neighbor* and *empty space embedding*. The former measures the distribution of active-to-active distances whereas the latter measures how uniformly the actives are embedded within the decoy space. Benchmarks are constructed by selecting a subset of the data that minimizes the difference between the two functions, ensuring the actives are maximally diverse and uniformly embedded within the decoy set.

A limitation of MUV unbiasing is that the measure ignores the distribution of inactive-to-inactive distances. Because machine learning algorithms do not care which class is mapped to actives and which is mapped to inactives, a machine learning system could simply memorize clusters among inactives to defeat a benchmark. Unintuitively, such clusters can be vast: because the algorithm merely needs to draw a hyperplane separating the actives from inactives, the data points do not need to be close together. That is, the cells in the Voronoi partitioning can be large. If it is the case that actives are closer to other actives and inactives are closer to other inactives, then looking up the label of the nearest data point will give the correct answer, as illustrated in Figure 1. This is true even if the absolute distances are high, as it is the relative distances which matter. All points that fall to one side of the decision boundary will be assigned to the same class, and the distance of the point to the boundary is irrelevant.

The technique of finding the label of the nearest data point is known as *one nearest neighbor* (1-NN). 1-NN has been previously used as a comparison baseline because of its weak predictive performance[20]. As a practical matter, benchmarks are often used to chose an algorithm from amongst a group of alternatives. Therefore, neither tests which are too easy (and passed by all considered systems) nor too difficult (and all systems fail) are useful. If 1-NN is sufficient to solve a benchmark, it is likely that the training and validation sets are too similar or that the test assumes access to unreasonably powerful discriminant features.

In this paper we formally define a bias measurement for ligand-based benchmarks that describes the ability of 1-NN to solve a validation set by memorizing the training data. We evaluate a variety of ligand-based benchmarks and demonstrate that our bias measurement is predictive of test accuracy for a wide range of machine learning algorithms, features, distance metrics, and unbiasing techniques. Our major result is that these observations are consistent with the view that every evaluated benchmark rewarding machine learning models for overfitting the training data, rather than the desired behavior of generalizing to novel data. While every benchmark we examined suffers from this bias, we further report which benchmarks are more and less resistant to being solved through overfitting their training data. We then experiment with a new bias reduction technique. We introduce an algorithm to minimize the redundancy bias, apply it to the analyzed benchmarks, and report classification performance over their native and unbiased forms. We provide the source code for both our bias measurement and reduction algorithms.

## Bias Measurement Method

We begin constructing our bias measure by extending the MUV data distribution functions from Rohrer and Baumann[17] to a machine learning setting comprising both training and test sets. Rohrer and Baumann define the *nearest neighbor* function as the proportion of actives for which the distance to their nearest neighbor active is less than a threshold. Rohrer and Baumann further define the *empty space embedding* function as the proportion of decoys for which the distance to their nearest active is less than a threshold. However, MUV only operates over a single set, which is to be used as the validation. This is an appropriate approach when the behavior of the analyzed method does not depend on training data, as with physics simulation. In contrast, any comparison of machine learning methods must define both the training and validation sets. Therefore, we add parameters to the MUV functions to specify which sets to analyze. We observe that providing sets of molecules as parameters allows us to use a single nearest neighbor function for both the MUV nearest neighbor and empty space embedding measures: let

$$S_{(V,T,d)} = \frac{1}{|V|} \sum_{v \in V} I_d(v, T)$$

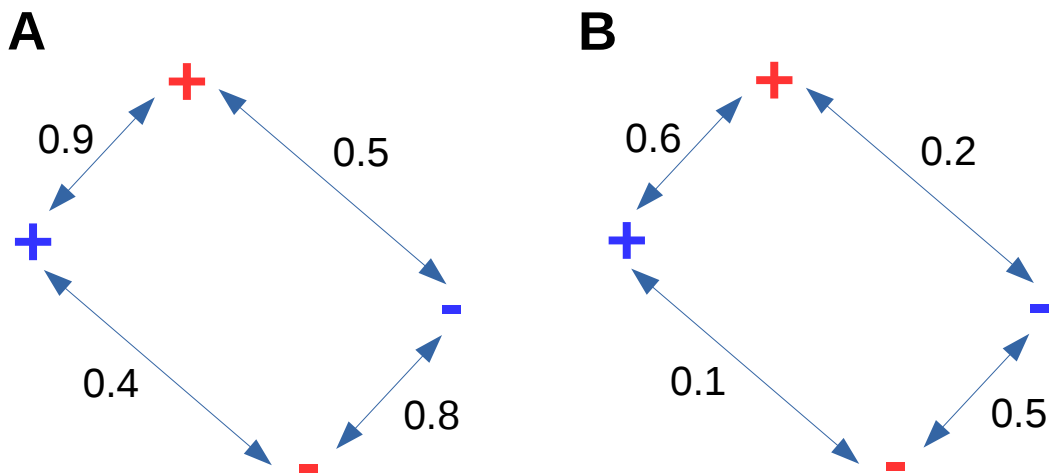

Figure 1: Relative distance analysis does not simply consider the domain of applicability. AVE bias shows that models may still achieve high accuracy, despite the validation set being dissimilar enough to fall outside the domain of applicability. We illustrate with two sets of data points. Each set consists of four molecules, where '+' signs indicate positive examples and '-' signed indicate negative examples. Blue and red colors indicate partitioning into training and validation sets, respectively. Black arrows indicate the similarities between pairs of molecules. Without loss of generality, suppose a similarity of 0.7 is considered necessary to have good predictions. On the left side (A), we enforce similarity between actives to be greater than 0.7 whereas on the right side (B) we enforce dissimilarity such that similarity between actives pairs are less than 0.7. The bias of 0.8 is the same for both sets ($B_{left} = (0.9 - 0.5) + (0.8 - 0.4) = B_{right} = (0.6 - 0.2) + (0.5 - 0.1)$). Predictive models should do well for the left dataset (A) because of its high training-validation similarity, while a "domain of applicability" argument suggests that predictive models should fail for the right dataset (B) because of low similarity between training and validation sets. However, our bias analysis predicts predictive models would do well for both datasets: because validation actives (inactives) are more similar to training actives (inactives) than to training inactives (actives), a machine learning system can draw a hyperplane separating positives from negatives.

be a nearest-neighbor function between two sets of molecules where $V$ is the set of validation molecules; $T$ is the set of training molecules; $d \in [0, 1]$ is a similarity distance threshold; and $I_d(v, T) = 1$ if the distance from validation molecule $v$ to its nearest neighboring molecule in $T$ is smaller than $d$, and 0 otherwise.

We define the cumulative nearest-neighboring function as:

$$H_{(V,T)} = \frac{1}{|D|} \sum_{d \in D} S(V, T, d), \quad D = \{0, 0.01, \cdots, 1\}$$

We define $V_a, V_i, T_a, T_i$ to be the sets of validation actives, validation inactives, training actives, and training inactives, respectively (The analysis holds for test sets, as well as validation, but we use "validation" throughout this paper for consistency.) $D$ is a non-empty set of distance thresholds defined over a distance metric with the range $[0-1]$. Then define the Asymmetric Validation Embedding (AVE) bias between validation and training sets as:

$$B_{(V_a, V_i, T_a, T_i)} = (H_{(V_a, T_a)} - H_{(V_a, T_i)}) + (H_{(V_i, T_i)} - H_{(V_i, T_a)})$$

where the closer $B$ gets to zero the less biased a benchmark becomes. For brevity, we define $H_{(V_a, T_a)}$ as $AA$, $H_{(V_a, T_i)}$ as $AI$, $H_{(V_i, T_i)}$ as $II$, and $H_{(V_i, T_a)}$ as $IA$. Then,

$$B_{(V_a, V_i, T_a, T_i)} = (AA - AI) + (II - IA)$$

This bias measurement is made of two parts. The left $(AA - AI)$ term is, as in MUV, a measure of how clumped the validation actives are among the training actives. The right $(II - IA)$ term extends the MUV analysis to also measure the degree of clumping among the inactives , because classification techniques can learn patterns in either class .

## Results

In the following we analyze the AVE bias of multiple different benchmarks prepared with different methodologies and data sources. Unless specified otherwise, for computing the AVE term, we use Tanimoto[21] (Jaccard) distance over the ECFP4 fingerprint[22]. We use four commonly used classification algorithms to demonstrate the effect of the bias on the assessment of performance: Random Forests (RF), Support Vector Machines (SVM), Logistic Regression (LR), and K-Nearest-Neighbors (KNN). All classification algorithms are implemented in the SciKit-Learn python library[23] and were used with default parameters with the exception of *RF(n_estimators=100)* and *KNN(k=1, metric='jaccard', algorithm='brute')*. With $k=1$, KNN always perfectly memorizes the training set; this provides an unambiguous baseline against which to compare predictive performance over held-out data, as the capacity of the model is necessarily sufficient to learn the training data. The other algorithms are similar: the median training AUC for all algorithms over all training sets was >0.99. To investigate the correlation between the AVE bias and the predictive performance of the models for each benchmark target, we compute Pearson ($\rho$) and Kendall ($\tau$) correlations along with the correlation coefficient ($r^2$) directly from the raw data using SciPy[24].

### J&J Benchmark

This benchmark consists of 1230 targets and approximately 2 million activities and was released by Unterthiner et al[1] as part of their work on multi-task deep neural networks. Activity data was gathered from ChEMBL[25] (version 17). Each data point was labeled as active if its measured bioactivity was lower than $1 \mu M$ or inactive if its measured bioactivity was higher than $30 \mu M$. The dataset ligands were clustered by single-linkage clustering algorithm[26] using Tanimoto[21] similarity and ECFP12[22] fingerprints (with unpublished similarity cutoff). Then, the activities were partitioned into 3-fold cross-validation based on a 3-fold partitioning of the ligand clusters. In the following, we used a subset of this bench-mark consisting of 560 targets for which there were at least 100 active compounds. The benchmark data can be downloaded from `http://www.bioinf.at/research/VirtScreen`.

Figure 2 illustrates the ranking performance achieved using the four classifiers and the biases measured for the datasets and the correlations between the two. The vast majority of datasets in this benchmark have a positive AVE bias, which implies that they may be trivially classified. Indeed, all four classifiers performed well (similar to the ones published by Unterthiner et al.[1]) where the RF, SVM, LR, and KNN models achieved average AUCs of 0.85, 0.81, 0.84, and 0.76, respectively. In order to assess the correlation between bias and rank-order performance we computed the Pearson ($\rho$) and Kendall ($\tau$) correlations as well as the correlation coefficient ($r^2$). For all classifiers, we observed high correlations between bias and AUC. For example, the RF model yielded correlations of $\rho = 0.84$, $\tau = 0.70$, and $r^2 = 0.71$. These correlations are higher than the correlation between RF AUC and either part of the AVE bias term. Figure S5 shows the ($AA - AI$) term alone generated correlations of $\rho = 0.74$, $\tau = 0.54$, and $r^2 = 0.55$, while Figure S6 shows the (II-IA) term alone generated correlations of $\rho = 0.47$, $\tau = 0.38$, and $r^2 = 0.22$.

These correlations between bias and AUC did not depend on the choice of fingerprints. In Figure 3, we show the same analysis performed both with MACCS fingerprints[27] and the simple fingerprints from the MUV paper[17]. For the MACCS fingerprints, the predictive performance remained high, with AUCs for RF of 0.82, SVM of 0.78, LR of 0.79, and 1-NN of 0.72. The correlation between bias and AUC were also high at $\rho = 0.82$, $\tau = 0.73$, and $r^2 = 0.68$ for the RF model. The simple fingerprints are mostly based on counting different atom-types ("count all atoms, heavy atoms, boron, bromine, carbon, chlorine, fluorine, iodine, nitrogen, oxygen, phosphorus, and sulfur atoms in each molecule") along with a few molecular properties ("the number of H-bond acceptors, H-bond donors, the logP, the number of chiral centers, and the number of ring systems")[17]. Simple fingerprints gave predictive performances that were quite close to more complicated features (with the exception of 1-NN): RF achieves an AUC of 0.78, SVM of 0.75, LR of 0.74, and 1-NN of 0.57. This impressive performance should be shocking, as simple fingerprints convey no information about the shape of the molecule or the relative position of the atoms. For any active compound, we could imagine many molecules that have the same number of carbons and various heteroatoms but distinct topologies, of which we should expect only few to bind[28]. Therefore, the robust performance of systems using atom-count based features may simply be explained by weaknesses in the test. Indeed, the correlation between bias and AUC is $\rho = 0.82$, $\tau = 0.69$, and $r^2 = 0.67$ for the RF model, and suggests the topology of the validation data explains much of the performance of the predictive systems.

**MUV Benchmark**

This benchmark consists of 17 targets with 30 actives and 15000 inactives per target[17]. Actives and inactive bioactivity measurements were selected from PubChem BioAssays[29]. In the construction of the benchmark, each target set is optimized to minimize the difference between the actives *nearest neighbor function*, which quantifies the similarity within the active set, and the *empty space function*, which quantifies how well the actives are embedded within the inactives space. We randomly partitioned each target set into 3-fold cross-validation sets. The benchmark data can be downloaded from `https://www.tu-braunschweig.de/pharmchem/forschung/baumann/muv`.

Figure 4 illustrates the correlations between AUC and AVE bias for the four classifiers. All classifiers show high correlations with $r^2$ ranging between 0.73 and 0.88. Because the MUV benchmark was optimized to minimize a function similar to our $AA - AI$ component of the AVE it is interesting to plot the $AA - AI$ and $II - IA$ components separately. We can see that the $AA - AI$ component is negative for all targets. This negativity suggests that validation actives are generally more similar to training inactives than to training actives and thus classification of actives is challenging for this set. However, the $II - IA$ bias component is positive and similar for all targets ($\sim 0.4$) which suggests that the classification of inactives is generally easy for this benchmark. We note that the MUV benchmark was optimized to minimize a function similar to $AA - AI$ with respect to molecular fingerprints much simpler than the ECFP4 used in our case. Evaluation of the bias using fingerprints similar to the ones used by Rohrer and Baumann resulted, as expected, with $AA - AI$ term close to zero (data not shown). Figures S8 and S9 show that the correlations between AUC and the full AVE bias are higher than the correlation between AUC and either part of the AVE bias term for every model except 1-NN.

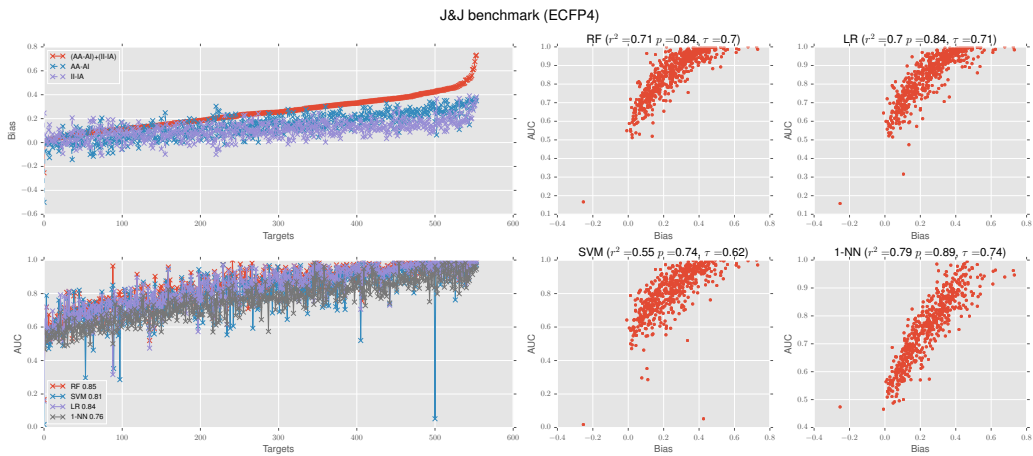

Figure 2: Illustration of the performance of four commonly used classification algorithms: Random-Forests (RF), Logistic Regression (LR), Support Vector Machines (SVM), and K-Nearest-Neighbors (KNN) over the J&J benchmark[1] using single-linkage similarity clustering. Each ML algorithm shows clear correlation between AVE bias and AUC (see text for discussion). Top left: calculated AVE bias for 560 benchmark targets with more than 100 active compounds. Bottom left: AUC performance of the different algorithms over the 560 targets ordered by their corresponding AVE bias values. Right: scatter plots of the correlation between AUC and AVE bias for each of the classification algorithm over the 560 targets.

To further investigate the distribution of active and inactive examples, we generated t-SNE[30] plots of the MUV datasets. The t-SNE technique embeds the 17-dimensional MUV simple descriptor data into a 2-dimensional projection while attempting to preserve the relative distances between data points in the original 17-dimensional space. We used the t-SNE implementation in the SciKit-Learn python library[23] with perplexity value of 30.0 and the 'jaccard' distance metric. Although different perplexity values yield different projections we noticed similar characteristics of active and decoy distributions (data not shown).

Figure 5 shows representative plots for 4 of the MUV targets. These projections show that actives often appear close to other actives, despite MUV being designed to select an even distribution of actives against the inactive background. The projections also depict a large amount of clustering among the decoys, which could permit good performance through memorization of the inactive class. Additionally, even when the decoys are not in dense clumps, they are often far from actives and close to other decoys. Therefore, if MUV data sets are randomly split into training and validation sets, the validation inactives are similar to training inactives and good performance can be achieved by learning patterns of negatives.

### ChEMBL time-split Benchmark

Time-split data is rare in the public domain. In order to approximate a time-split benchmark we constructed a training set of 42 targets from version 19 of the ChEMBL bioactivity database[25] and a validation set of their corresponding bioactivities that were first introduced in version 20. We note however that the first appearance of data in a high ChEMBL version does not guarantee that the data was reported after data in previous ChEMBL versions.

Figure 6 illustrates the correlations between bias and AUC. We observe clear correlation between bias and AUC for all learning methods investigated, ranging from 0.42 to 0.64 for the Kendall rank-order correlation. As a specific example, the RF model yielded $\rho = 0.84$, $\tau = 0.64$, and $r^2 = 0.7$. These correlations are considerably higher than the correlation between RF AUC and either part of the AVE bias term. Figure S15 shows the $(AA - AI)$ term alone generated correlations of $\rho = 0.26$, $\tau = 0.14$, and $r^2 = 0.07$ with RF AUC, while Figure S16 shows the (II-IA) term alone generated correlations of $\rho = 0.6$, $\tau = 0.45$, and $r^2 = 0.36$.

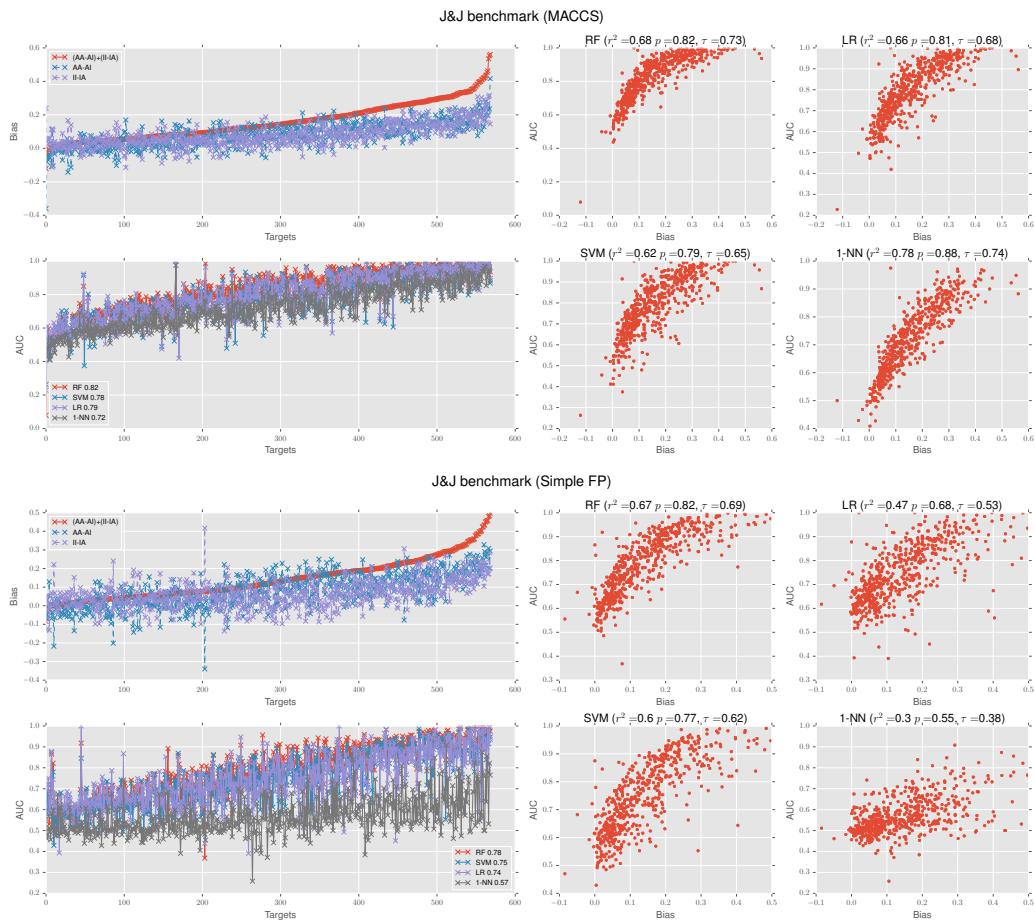

Figure 3: Evaluation of the performance of alternative fingerprints over the J&J benchmark. In both the MACCS (top) and MUV-simple (bottom) fingerprints we notice clear correlations between AVE bias and AUC values, regardless of the classification algorithm being used.

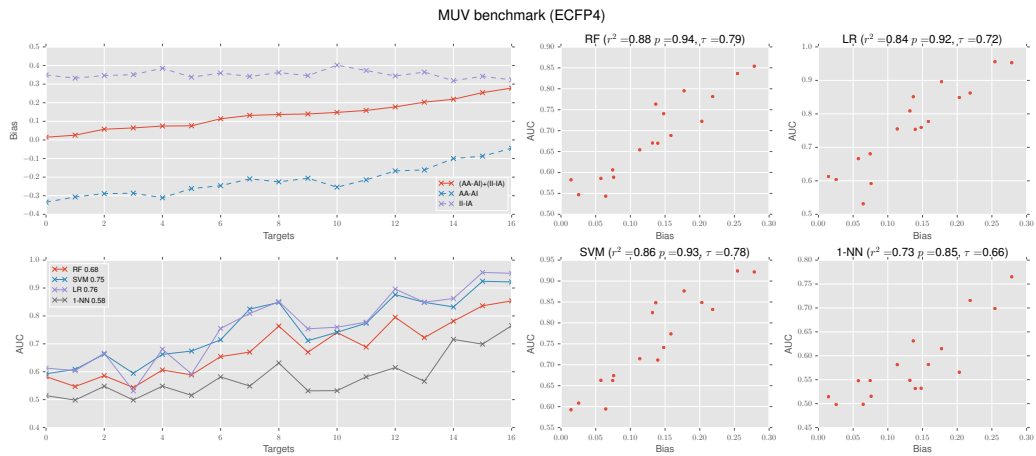

Figure 4: Illustration of the performance of four commonly used classification algorithms: Random-Forests (RF), Logistic Regression (LR), Support Vector Machines (SVM), and K-Nearest-Neighbors (KNN) over the MUV benchmark [17] using random partitioning. Each ML algorithm shows clear correlation between AVE bias and AUC (see text for discussion). Top left: calculated AVE bias for 17 benchmark targets. Total AVE bias, AVE bias from active similarities, and AVE bias from inactive similarities are shown separately. Bottom left: AUC performance of the different algorithms over the 17 targets ordered by their corresponding AVE bias values. Right: scatter plots of the correlation between AUC and AVE bias for each of the classification algorithm over the 17 targets.

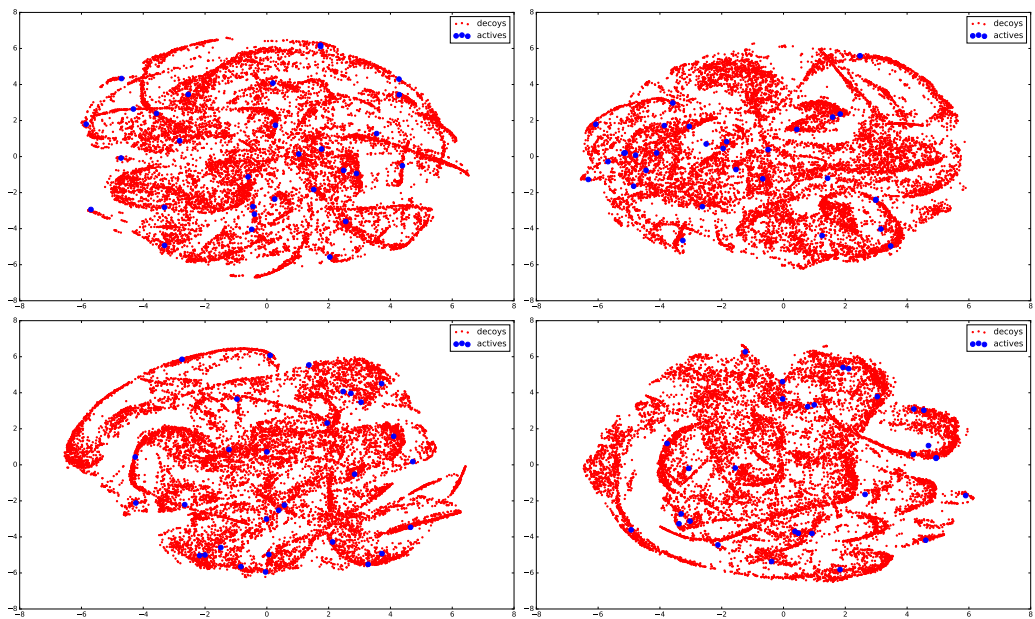

Figure 5: t-SNE plot of 4 representative MUV datasets using the MUV-simple fingerprints and Tanimoto distance. Using the t-SNE algorithm, we reduce the dimensionality of every data point from the 17 dimensions of the MUV-simple fingerprint to 2 dimensions while attempting to preserve the high-order distances in the lower dimensions.

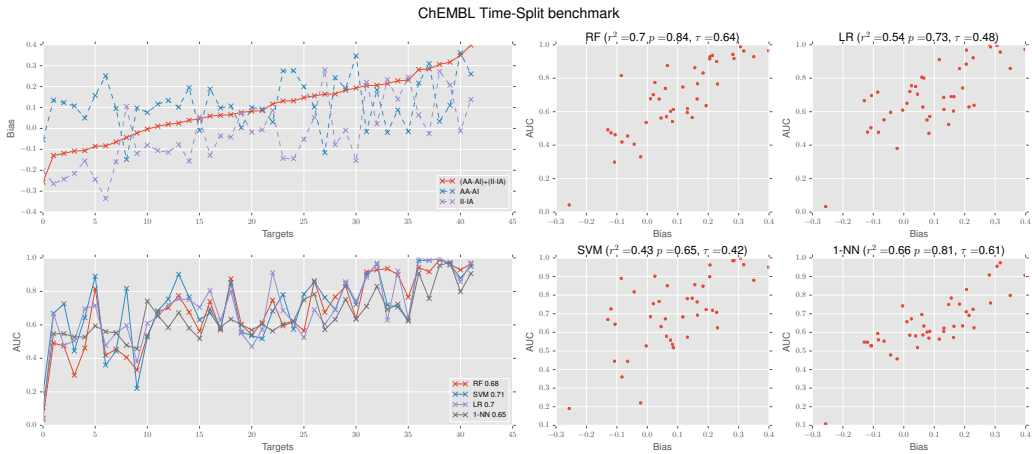

Figure 6: Illustration of the performance of four commonly used classification algorithms: Random-Forests (RF), Logistic Regression (LR), Support Vector Machines (SVM), and K-Nearest-Neighbors (KNN) over a time-split approximation benchmark employing ChEMBL[25] version 19 and 20 as training and validation sets, respectively. Each ML algorithm shows clear correlation between AVE bias and AUC (see text for discussion). Top left: calculated AVE bias for 42 benchmark targets. Bottom left: AUC performance of the different algorithms over the 42 targets ordered by their corresponding AVE bias values. Right: scatter plots of the correlation between AUC and AVE bias for each of the classification algorithm over the 42 targets.

## Tox-21 2014 Challenge Benchmark

This benchmark was released as part of the Tox21 2014 challenge[31] (`https://tripod.nih.gov/tox21/challenge`). It consists of 12 target systems each of which includes about 200-1000 active and 6000-9000 inactive molecules. The corresponding validation sets include 4-48 active and 186-288 inactive molecules. In our analysis we excluded two targets for which there were less than 10 active molecules in the validation sets. The benchmark data can be download from `https://tripod.nih.gov/tox21/challenge/data.jsp`.

Figure 7 illustrates clear correlation between bias and AUC of 0.41 to 0.7, 0.64 to 0.83, and 0.42 to 0.56, for $r^2$, Pearson, and Kendall correlations, respectively. This is typically but not always better than the correlation with the $(AA - AI)$ term alone, as shown in Figure S17.

## PCBA Benchmark

The PubChem BioAssays benchmark uses data on over 400,000 molecules measured by high-throughput screens[29]. A subset of 128 targets has been used in previous evaluations for modern machine learning methods[2;32]. Originally, the training and validation data were split randomly, which lead to significant similarity between training and validation[2]. More recent versions have attempted to generate challenging data partitions by clustering the training data on Bemis-Murcko scaffolds[33] and assigning entire clusters into training or validation sets[34].

We analyzed the correlation between bias and predictive AUC performance for both of these versions. In Figure 8, we present the correlation for the randomly-partitioned version of the

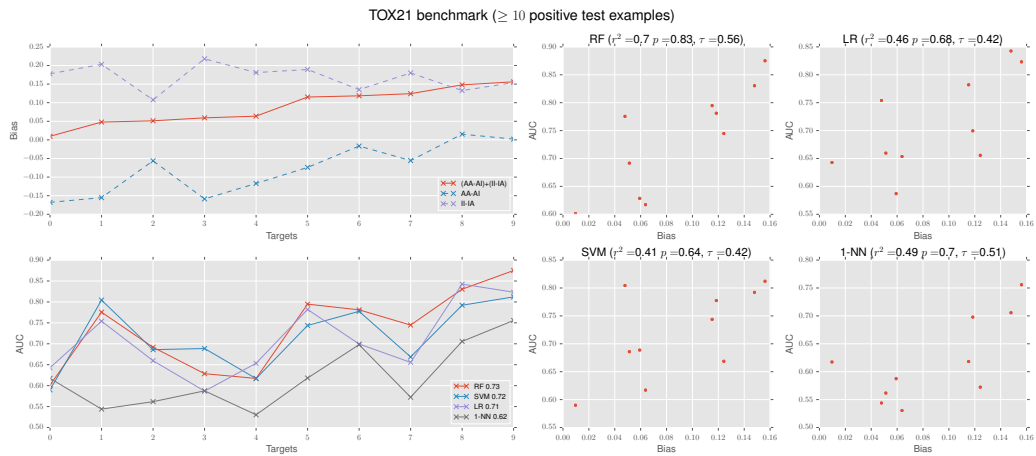

-AI

Figure 7: Illustration of the performance of four commonly used classification algorithms: Random-Forests (RF), Logistic Regression (LR), Support Vector Machines (SVM), and K-Nearest-Neighbors (KNN) over the Tox-21 2014 benchmark[31]. Each ML algorithm shows substantial correlation between AVE bias and AUC (see text for discussion). Top left: calculated AVE bias for 10 benchmark targets. Bottom left: AUC performance of the different algorithms over the 10 targets ordered by their corresponding AVE bias values. Right: scatter plots of the correlation between AUC and AVE bias for each of the classification algorithm over the 10 targets.

PCBA benchmark. All machine learning approaches performed well, with RF achieving an AUC of 0.84, SVM 0.82, LR 0.84, and 1-NN 0.72. Much of this performance can be attributed to the benchmark bias, as the $r^2$ for the models varied from 0.55 to 0.7, Kendall's $\tau$ varied from 0.58 to 0.66, and Pearson's $\rho$ varied from 0.74 to 0.84. For LR and SVM, these correlations are higher than the correlation between AUC and the $(AA - AI)$ part of the AVE bias term, as shown in Figure S20 and Figure S21. For 1-NN and for RF $r^2$ and $\rho$, however, the correlation was higher with $(AA - AI)$.

In Figure 9, we present the correlation for the Murcko-partitioned version of the PCBA benchmark. We note that we implemented our own partitioning of the data using generic-Murcko scaffolds, rather than the non-generic partitioning provided by MoleculeNet[34]. The Murcko scaffold partitioning showed small reductions in both median bias (from over 0.2 to under 0.15) and mean AUC, with RF achieving an AUC of 0.81, SVM 0.79, LR 0.81, and 1-NN 0.67. However, the correlation remained quite strong, as the $r^2$ for the models varied from 0.47 to 0.66, Kendall's $\tau$ varied from 0.52 to 0.64, and Pearson's $\rho$ varied from 0.68 to 0.81.

**ToxCast Benchmark**

The ToxCast benchmark comprises 8615 molecules measured by 617 phenotypic assays, including protein, cell-based, and animal developmental measures, performed by the EPA[35]. The benchmark is available for download from MoleculeNet[34].

As with the PubChem BioAssay benchmark, we investigate the bias-AUC performance correlations of the benchmark using random and scaffold-based partitioning In Figure 10, we present the correlation for the randomly-partitioned version of the ToxCast benchmark. ToxCast proved to be one of the more challenging benchmarks, as average AUCs ranged from 0.58 for 1-NN to 0.67 for RF. However, substantial correlations remained, with the $r^2$ for the models varying from 0.38 to 0.66, Kendall's $\tau$ varying from 0.55 to 0.65, and Pearson's $\rho$ varying from 0.61 to 0.81. In each case, the lowest correlation was due to SVM, which performed poorly on some targets at all levels of bias. The results for the generic-

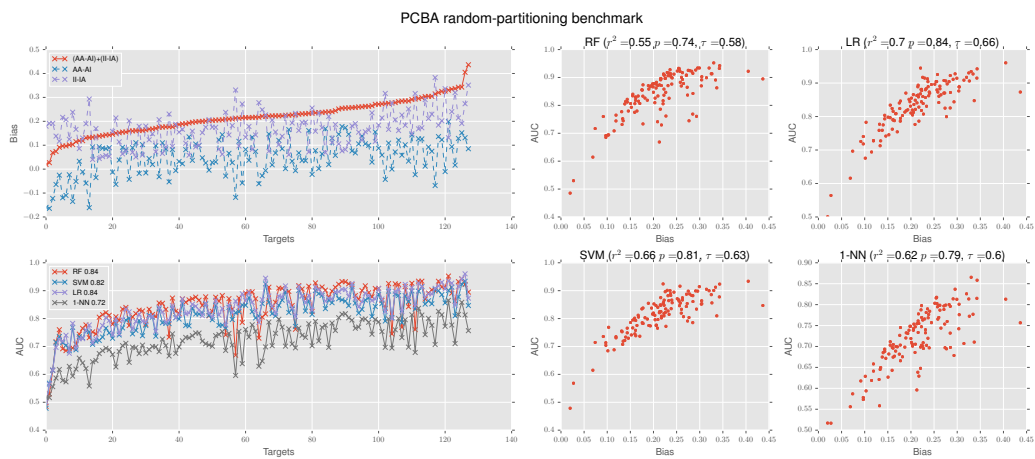

Figure 8: Illustration of the performance of four commonly used classification algorithms: Random-Forests (RF), Logistic Regression (LR), Support Vector Machines (SVM), and K-Nearest-Neighbors (KNN) over the PCBA benchmark[2;34] using random partition cross-validation. Each ML algorithm shows visible correlation between AVE bias and AUC (see text for discussion). Top left: calculated AVE bias for 128 benchmark targets. Bottom left: AUC performance of the different algorithms over the 128 targets ordered by their corresponding AVE bias values. Right: scatter plots of the correlation between AUC and AVE bias for each of the classification algorithm over the 128 targets.

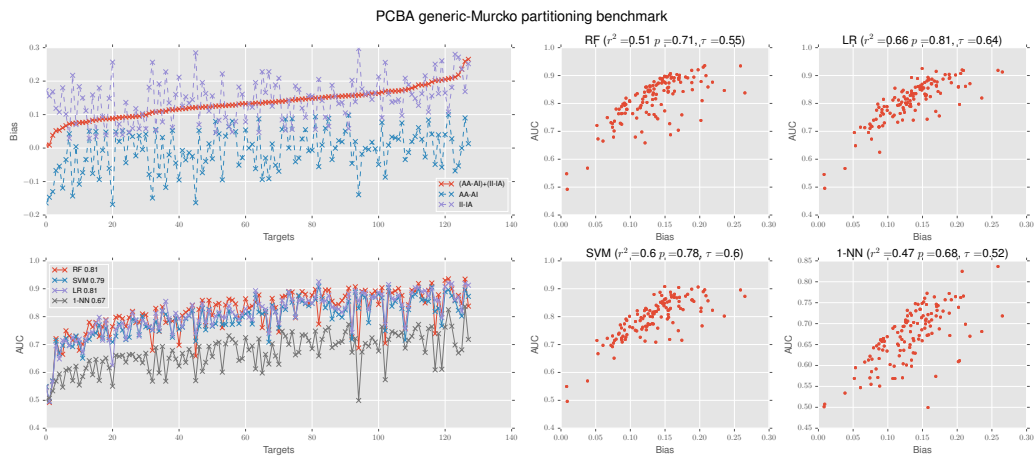

Figure 9: Illustration of the performance of four commonly used classification algorithms: Random-Forests (RF), Logistic Regression (LR), Support Vector Machines (SVM), and K-Nearest-Neighbors (KNN) over the PCBA benchmark[2;34] using generic-Murcko partition cross validation. Each ML algorithm shows clear correlation between AVE bias and AUC (see text for discussion). Top left: calculated AVE bias for 128 benchmark targets. Bottom left: AUC performance of the different algorithms over the 128 targets ordered by their corresponding AVE bias values. Right: scatter plots of the correlation between AUC and AVE bias for each of the classification algorithm over the 128 targets.

Murcko partitioned version of the ToxCast benchmark are shown in Figure 11. Under this partitioning, we measured an $r^2$ for the models from 0.29 to 0.56, Kendall's $\tau$ varied from 0.43 to 0.56, and Pearson's $\rho$ varied from 0.53 to 0.75. Figures S25,S26,S27, and S28 show the correlation between AUC and either part of the AVE bias. The AUC correlation is higher with the full AVE term for every ML algorithm under every correlation measure.

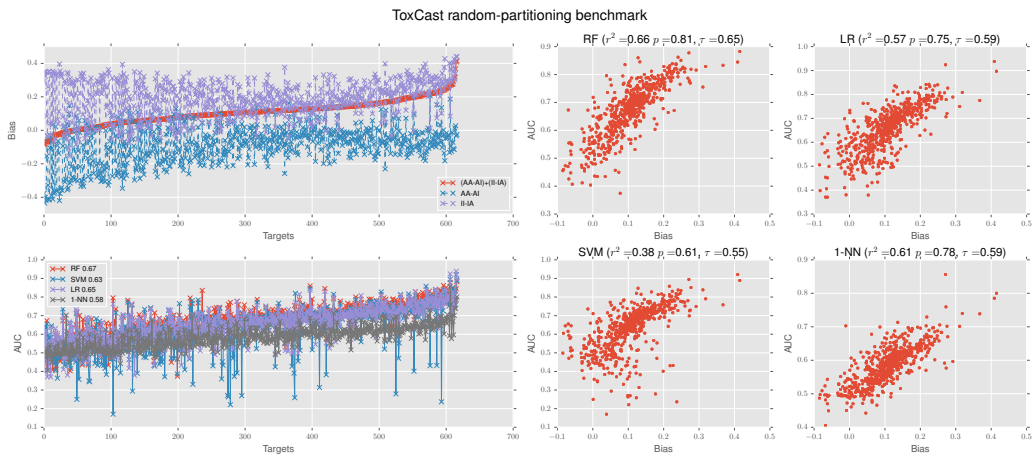

Figure 10: Illustration of the performance of four commonly used classification algorithms: Random-Forests (RF), Logistic Regression (LR), Support Vector Machines (SVM), and K-Nearest-Neighbors (KNN) over the ToxCast benchmark[34] using random partition cross-validation. Each ML algorithm shows clear correlation between AVE bias and AUC (see text for discussion). Top left: calculated AVE bias for phenotypic conditions. Bottom left: AUC performance of the different algorithms over the 617 phenotypes ordered by their corresponding AVE bias values. Right: scatter plots of the correlation between AUC and AVE bias for each of the classification algorithm over the 617 targets.

**Sider Benchmark**

The Sider database catalogues adverse drug reactions (ADR) associated with marketed drugs in public documents and package inserts[36]. We analyzed MoleculeNet[34] version of Sider, which grouped the adverse reactions of 1427 drugs into 27 ADR classes. Random partitioning, depicted in Figure 12, shows $r^2$ correlations of 0.57 to 0.82, $\rho$ of 0.75 to 0.91, and $\tau$ of 0.58 to 0.67. Generic-Murcko partitioning, depicted in Figure 13, shows $r^2$ correlations of 0.5 to 0.64, $\rho$ of 0.71 to 0.8, and $\tau$ of 0.45 to 0.66. Figures S30,S31,S32, and S33 show the correlation between AUC and either part of the AVE bias. The AUC correlation is higher with the full AVE term for every ML algorithm under every correlation measure.

**Proprietary Merck Benchmark**

We investigated how well machine learning models that were trained on public data would transfer to predicting internal validation data generated by Merck. For each of 21 targets of interest to Merck (approximately half kinase and half non-kinase), we built machine learning models using training data from ChEMBL v20. The validation actives were taken from Merck internal data for each target. Actives were eliminated from the validation set if they had a Dice similarity for atom-pair feature-pair fingerprints[37;38] of more than 0.7 to any compound in ChEMBL, to enforce a lack of redundancy between training and validation sets. This dissimilarity was considered sufficient to eliminate training actives from the validation set. For each active, 40 validation inactives were chosen from ChEMBL v20; 20 inactives were chosen at random and 20 inactives were chosen as property-matched decoys[19]. We were blinded to the validation labels until after the predictions were delivered.

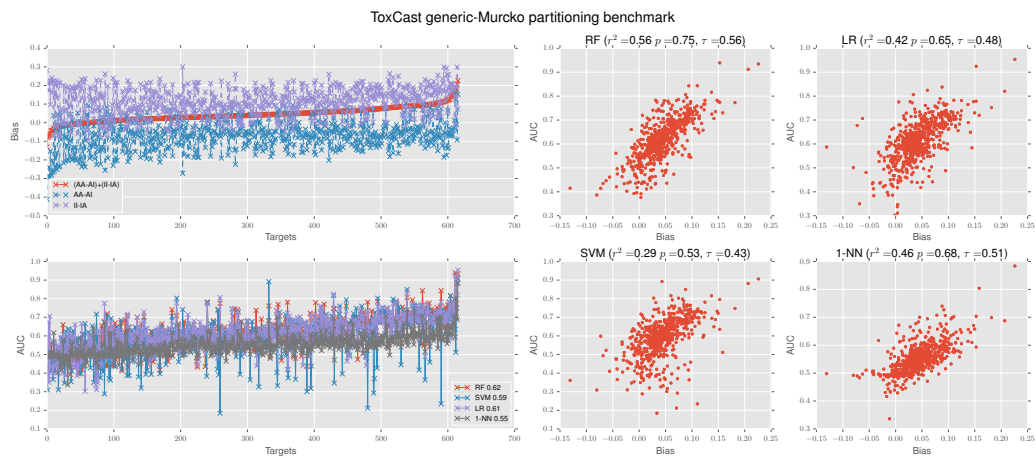

Figure 11: Illustration of the performance of four commonly used classification algorithms: Random-Forests (RF), Logistic Regression (LR), Support Vector Machines (SVM), and K-Nearest-Neighbors (KNN) over the ToxCast benchmark[34] using generic-Murcko partition cross-validation. Each ML algorithm shows correlation between AVE bias and AUC, although less than with random partitioning (see text for discussion). Top left: calculated AVE bias for 617 phenotypic conditions. Bottom left: AUC performance of the different algorithms over the 617 phenotypes ordered by their corresponding AVE bias values. Right: scatter plots of the correlation between AUC and AVE bias for each of the classification algorithm over the 617 targets.

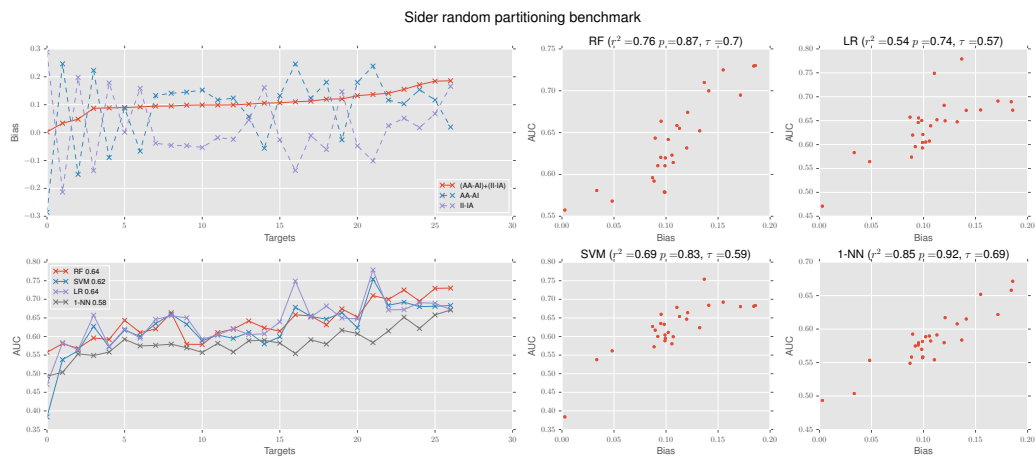

Figure 12: Illustration of the performance of four commonly used classification algorithms: Random-Forests (RF), Logistic Regression (LR), Support Vector Machines (SVM), and K-Nearest-Neighbors (KNN) over the Sider benchmark[34;36] using random partition cross-validation. Each ML algorithm shows clear correlation between AVE bias and AUC (see text for discussion). Top left: calculated AVE bias for 617 benchmark targets. Bottom left: AUC performance of the different algorithms over the 27 ADR classes ordered by their corresponding AVE bias values. Right: scatter plots of the correlation between AUC and AVE bias for each of the classification algorithm over the 27 targets.

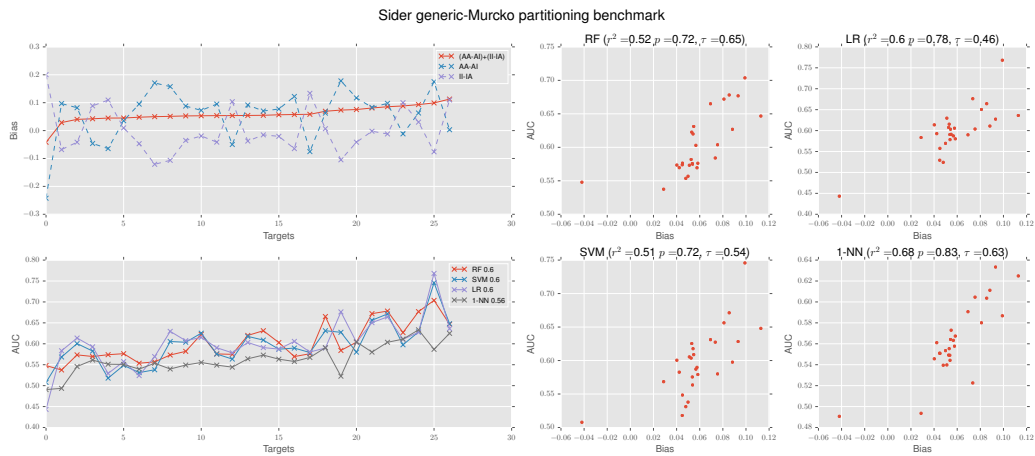

Figure 13: Illustration of the performance of four commonly used classification algorithms: Random-Forests (RF), Logistic Regression (LR), Support Vector Machines (SVM), and K-Nearest-Neighbors (KNN) over the Sider benchmark[34;36] using generic-Murcko partition cross-validation. Each ML algorithm shows clear correlation between AVE bias and AUC (see text for discussion). Top left: calculated AVE bias for 27 benchmark targets. Bottom left: AUC performance of the different algorithms over the 27 ADR classes ordered by their corresponding AVE bias values. Right: scatter plots of the correlation between AUC and AVE bias for each of the classification algorithm over the 27 targets.

Figure 14 shows the AVE bias analysis for this benchmark. Despite enforcing a fingerprint dissimilarity between training and validation actives, we find robust correlations between the measured AVE bias and model AUC. The measured $r^2$ ranged from 0.62 for RF to 0.81 for 1-NN, the $\rho$ ranged from 0.79 for SVM to 0.9 for 1-NN, and $\tau$ ranged from 0.66 for SVM to 0.74 for RF and 1-NN.

We then investigated whether it would be possible to make correct predictions without using positive data points; success would demonstrate that information was being carried by the negatives. We made predictions without seeing any active molecules by shuffling the model training data, as follows. For each pair of 10 unblinded targets, $X$ and $Y$, we took the actives from target $X$. We then eliminated every active that was more than 0.7 Dice similarity to any active for target $Y$. Note that this was the same threshold that was considered sufficient to exclude molecules too similar to training actives from the validation sets in the original benchmark. That is, every remaining active from $X$ could have been used as an inactive for $Y$ in the original test. We then added to the training set the inactives from $Y$. Therefore, every data point in the resulting training set for $Y$ could have been used as an inactive for $Y$ in the original test. We built a model using the resulting training set and evaluated the performance on target $Y$'s validation set. For example, we trained a model with BACE1 data and used it to identify ZAP70 actives and inactives, even though these proteins are unrelated and should not share ligands.

The expected behavior of such predictive models is to be random. However, as shown in Figure 15, every model achieves a modest average improvement over random performance. Over the 90 prediction pairs (training data from one of 10 unblinded targets, validation data from each of the other 9 targets), RF achieves an average AUC of 0.59, SVM of 0.6, LR of 0.56, and 1-NN of 0.52. More concerningly, the amount of AVE bias identifies which prediction pairs were difficult and which were easy. When the bias was highest, every algorithm except 1-NN could achieve an AUC higher than 0.9.

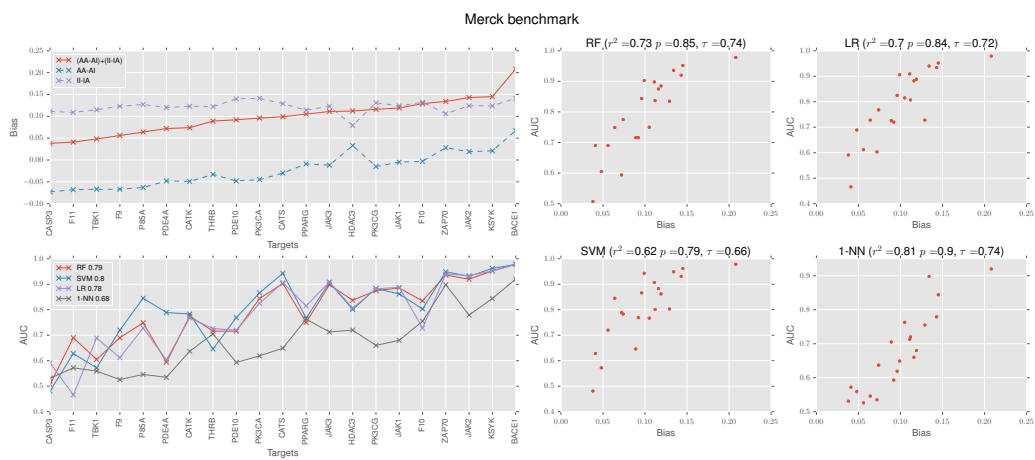

Figure 14: Illustration of the performance of four commonly used classification algorithms: Random-Forests (RF), Logistic Regression (LR), Support Vector Machines (SVM), and K-Nearest-Neighbors (KNN) over the proprietary Merck benchmark. Top left: calculated AVE bias for 21 benchmark targets. Each ML algorithm shows clear correlation between AVE bias and AUC (see text for discussion). Bottom left: AUC performance of the different algorithms over the 21 targets ordered by their corresponding AVE bias values. Right: scatter plots of the correlation between AUC and AVE bias for each of the classification algorithm over the 21 targets.

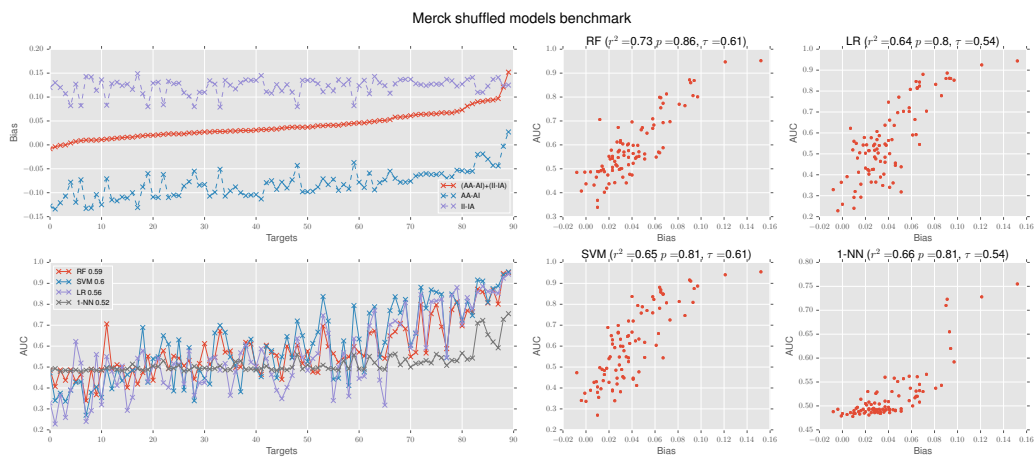

Figure 15: Illustration of the performance of four commonly used classification algorithms: Random-Forests (RF), Logistic Regression (LR), Support Vector Machines (SVM), and K-Nearest-Neighbors (KNN) over the proprietary Merck benchmark. Top left: calculated AVE bias for 90 pairs of different targets where the models are built using the training data of one target and predictions are made over the validation data of another target while maintaining less than 0.7 APFP-Dice similarity between the sets. Each ML algorithm shows clear correlation between AVE bias and AUC, although 1-NN struggles with low bias tests cases (see text for discussion). Bottom left: AUC performance of the different algorithms over the 90 target-pairs ordered by their corresponding AVE bias values. Right: scatter plots of the correlation between AUC and AVE bias for each of the classification algorithm over the 90 target-pairs.

## Discussion

Machine Learning is the practice of minimizing an error function over a set of evaluation data. However, the choice of correct error function and appropriate evaluation data remains in the hands of the domain expert. Incorrect choices during evaluation can lead to an arbitrarily bad gap between expected and delivered performance. It is the responsibility of the domain expert to ensure that the test accurately reflects both the character and the difficulty of the intended real world application.

The need to make predictions for examples that are *unlike* our training data define the character and the difficulty of cheminformatics. That is, the process of biochemical discovery pulls from nonstationary distributions, which are particularly challenging domains for machine learning. New synthetic techniques and newly-economically-accessible starting materials enable the construction of novel chemical matter. Unexplored classes of biological targets (such as protein-protein, protein-lipid, or protein-carbohydrate interactions) require new chemical structures[39]. In pharmaceutical discovery, patent law incentivizes exploration away from exemplified chemical space[40].

Furthermore, beyond the algorithmic difficulties of modeling nonstationary distributions, there are important challenges due the role of cheminformatics within the drug discovery process. It is rare that computational chemists solely drive a research program. Our recommendations are mediated through experimentalists so, for a computational technique to be useful, the predictions must change the mind of the other project chemists. In the limit, if a medicinal chemist has a fixed research agenda comprising a set of molecules already committed to be made and a set of molecules already sworn to be avoided, then prediction on these sets provides no benefit to the research program. Even perfect predictive accuracy would be redundant[2]. Computational predictions are graded by accuracy but weighted by the amount of experimentally-validated novelty that they generate; activity cliffs in matched molecular pairs[41;42] and molecular pairs with high 3D similarity and low 2D similarity[43] illustrate the utility of finding highly novel molecules. Indeed, drug discovery project teams empirically select outlier molecules for follow-up out of virtual screens, rather than close variants to known compounds[44]. Therefore, we must focus our assessments not on how well our models interpolate within a set of small changes to a scaffold but on how well they extrapolate to surprising or unknown molecules.

To this end, a number of techniques are used to reduce data redundancy and mitigate bias, such as scaffold clustering, fingerprint-based clustering, temporal splitting, and maximum unbiased validation. These are applied to avoid rewarding algorithms for good performance on trivial variations of training examples. However, these techniques focus on the similarity between actives, assume inactives are sufficiently diverse, ignore the effect of inactive-inactive similarity, and consider absolute rather than relative similarities. Figure 16 illustrates that only considering active-active distances, without comparing that distance relative to the inactive-inactive distances, can leave large numbers of trivial test cases: even if we use clustering to exclude test cases in the circle around point `b`, the rest of the points in the Voronoi cell are still much closer to `b` than any other training example. A simple lookup of the closest neighbor correctly labels every remaining example in the Voronoi cell, even though they seem to be sufficiently distant to be challenging tests.

As shown in Table 1, we investigated seven widely-used benchmarks that had been previously unbiased in various ways, yet each ligand-based benchmarks we investigated showed a substantial $r^2$ correlation coefficient between AUC and the AVE bias measurement. As discussed in the Results, the correlations between AUC and AVE bias are typically higher than between AUC and either piece of the AVE measure. Therefore, we propose to add AVE bias analysis to the cheminformatic armamentarium. In the same way that one might assess the robustness of a benchmark to clustering or MUV unbiasing, one can use our approach to assess the impact of redundancies among the inactives. For example, good performance on high bias test cases and poor performance on low bias test cases is consistent with overfitting to the redundancies between the training and validation sets. This scenario occurred for every algorithm we tried on every benchmark we analyzed. The likely consequence is that models will not generalize to prospective predictions, despite good benchmark performance.

---

[2]See comments on http://blogs.sciencemag.org/pipeline/archives/category/in-silico for many examples where computational contribution is limited to retrospective explanations.

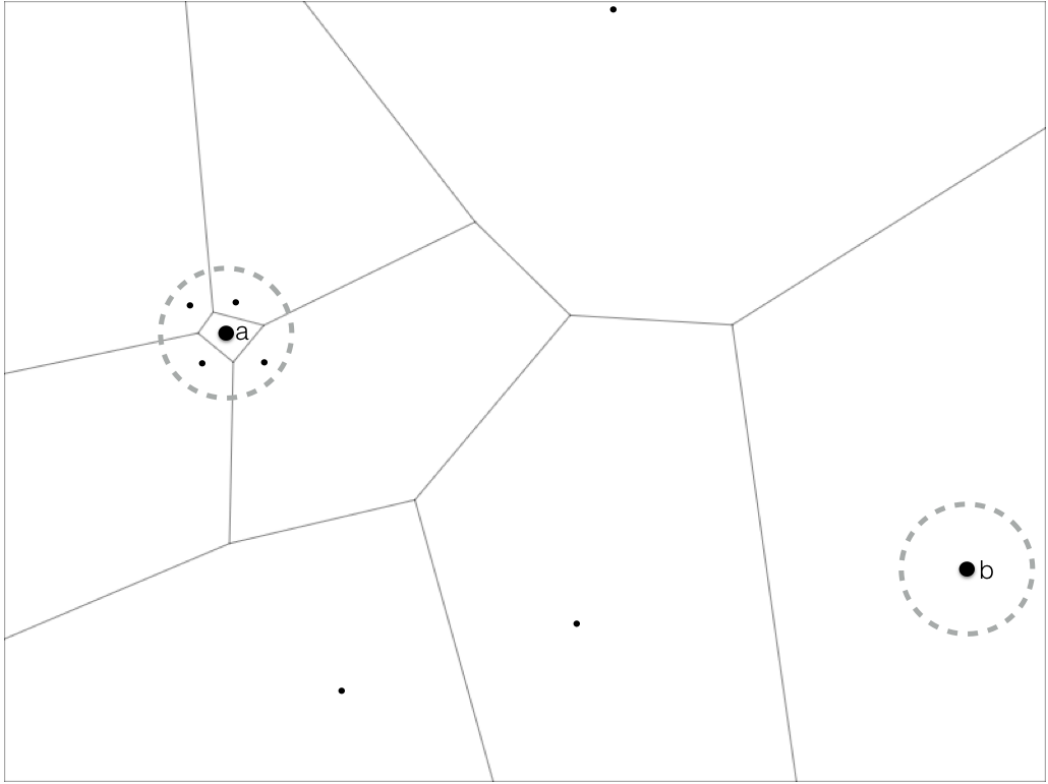

Figure 16: A fixed distance from an example can be too large or too small; consequently, filters based on fixed distances can remove too much or too little data. The dashed circle around point a encompasses the entirety of a's Voronoi cell along with examples from neighboring cells. Excluding all examples in the circle would unnecessarily lose data by eliminating difficult-to-classify data points on the cell border. On the other hand, excluding the dashed circle around point b leaves a large region of b's Voronoi cell, and points found in the region outside the circle and inside the cell can be labelled trivially through their relative similarity to b, despite being far away from b. In this case, excluding points only within the dashed circle does little to avoid points that are "too close" to b.

| Benchmark | Previous Unbiasing | $r^2$ | | | |
|---|---|---|---|---|---|
| | | RF | LR | SVM | 1-NN |
| J&J (ECFP) | Clustering | 0.71 | 0.7 | 0.55 | 0.79 |
| J&J (MACCS) | Clustering | 0.68 | 0.66 | 0.62 | 0.78 |
| J&J (SimpleFP) | Clustering | 0.67 | 0.47 | 0.6 | 0.3 |
| MUV | MUV | 0.88 | 0.84 | 0.86 | 0.73 |
| ChEMBL time-split | Approx time split | 0.7 | 0.54 | 0.43 | 0.66 |
| Tox21 | Random | 0.7 | 0.46 | 0.41 | 0.49 |
| PCBA | Random | 0.55 | 0.7 | 0.66 | 0.62 |
| PCBA | Murcko | 0.51 | 0.66 | 0.6 | 0.47 |
| ToxCast | Random | 0.66 | 0.57 | 0.38 | 0.61 |
| ToxCast | Murcko | 0.56 | 0.42 | 0.29 | 0.46 |
| SIDER | Random | 0.73 | 0.57 | 0.7 | 0.82 |
| SIDER | Murcko | 0.51 | 0.63 | 0.5 | 0.64 |
| Merck internal | Dice dissimilarity | 0.73 | 0.7 | 0.62 | 0.81 |
| Merck shuffled internal | | 0.73 | 0.64 | 0.65 | 0.66 |
| Merck kinase-kinase | Different targets + Dice dissimilarity | 0.63 | 0.51 | 0.39 | 0.66 |
| Merck kinase-nonkinase | | 0.47 | 0.41 | 0.47 | 0.21 |

Table 1: Summary of the $r^2$ correlation between AVE bias and the performance of different Machine Learning algorithms over the benchmarks discussed in this manuscript, and the Merck internal data. We find substantial correlations between AVE bias and AUC for every ML algorithm over every benchmark, and in spite of previous unbiasing of the data.

**Proprietary Merck benchmark discussion**

The results for a particularly extreme example are shown in proprietary Merck Benchmark Section above. In the shuffled benchmark, we built a predictive model for one target using another target's data. That is, we built a predictive model without showing a single active for the intended target. In fact, in the original benchmark design, every training active could have been chosen as an inactive for the predicted target. As better-than-random-prediction was achieved for many experiments, and no actives had been seen by the models, we conclude that information must be carried on training inactives. Worse, the amount of AVE bias could be used both to identify on which tests the models would perform well and on which they would fail. Despite the fact that the shuffling should have left no signal in the training data, the RF model, for example, achieves an $r^2$ of 0.73, Pearson's $\rho$ of 0.86, and Kendall's $\tau$ of 0.61.

We also controlled for the possibility that actives that were annotated against one target could correctly be informative in building models for related targets. Specifically, we split the targets into kinases and non-kinases; we then repeated our analysis by training on kinases and reporting predictive performance separately for kinases and non-kinases. These results are shown in Figures 17 and 18. As expected, kinase training data generalized to other kinases. As shown in Figure 18, AVE bias remained predictive of model performance for the non-kinase targets. Despite the protein targets being unrelated and no actives having been seen during training, tests with high AVE bias were solvable. Even when using modern machine learning methods, it is implausible that such benchmark performance will generalize prospectively.

**Temporal splitting**

It is tempting to think of temporal splits as a natural model of the pharmaceutical discovery process. However, temporal splitting only captures data about the molecules that were tested, which is not all molecules that were considered. Molecules that were rejected by the medicinal chemist for synthesis or assay do not appear in the temporal data at all; we do not get to see if a predictive model would make the same choices. In prospective uses, the predictive model would need to consider molecules that have not been pre-filtered by human judgement and prejudices. Therefore, a temporally-split validation set does not exactly model the same task as a discovery project.

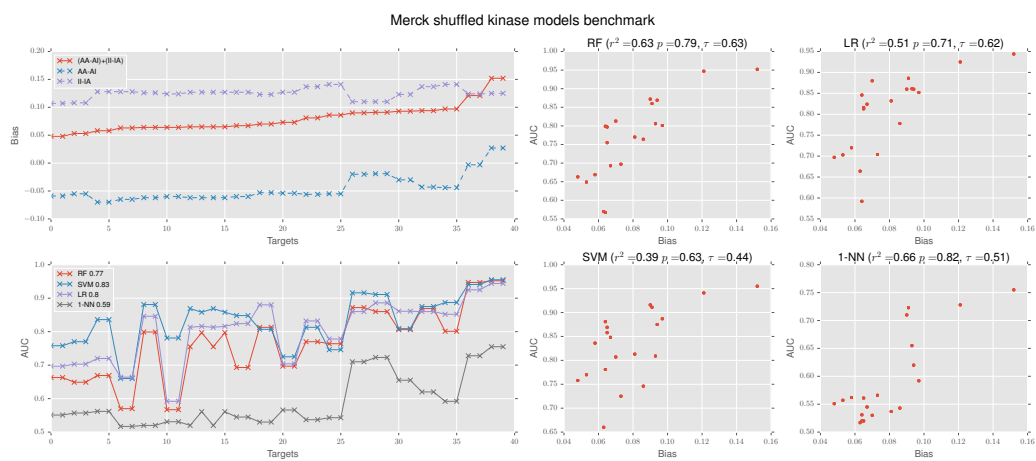

Figure 17: Illustration of the performance of four commonly used classification algorithms: Random-Forests (RF), Logistic Regression (LR), Support Vector Machines (SVM), and K-Nearest-Neighbors (KNN) over the proprietary Merck benchmark. Top left: calculated AVE bias for 50 pairs of different targets where the models are built using the training data of one kinase and predictions are made over the validation data of another kinase target, while maintaining less than 0.7 APFP-Dice similarity between the sets. Bottom left: AUC performance of the different algorithms over the 50 target-pairs ordered by their corresponding AVE bias values. Right: scatter plots of the correlation between AUC and AVE bias for each of the classification algorithm over the 50 kinase-kinase pairs.

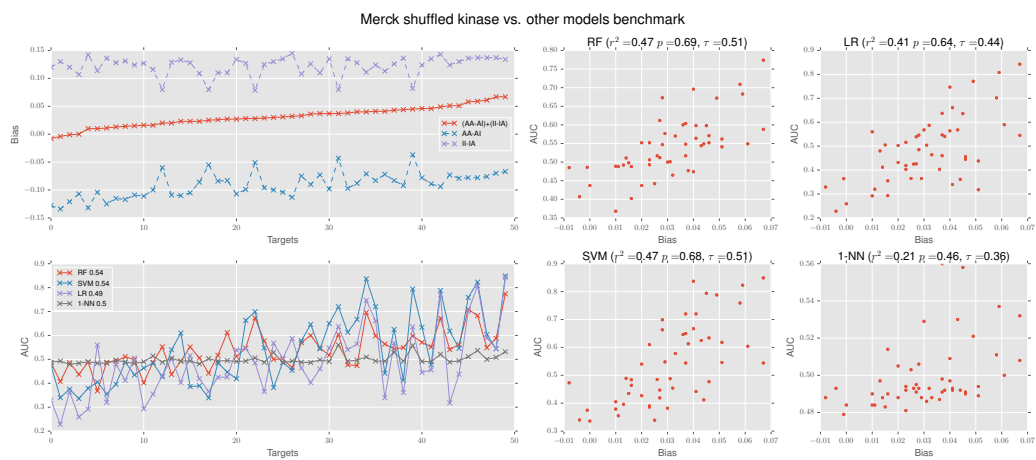

Figure 18: Illustration of the performance of four commonly used classification algorithms: Random-Forests (RF), Logistic Regression (LR), Support Vector Machines (SVM), and K-Nearest-Neighbors (KNN) over the proprietary Merck benchmark. Top left: calculated AVE bias for 50 pairs of kinase and non-kinase targets where the models are built using the training data of one kinase and predictions are made over the validation data of another non-kinase target, while maintaining less than 0.7 APFP-Dice similarity between the sets. When the bias is high, a model trained for unrelated proteins can still miraculously separate active and inactive molecules. Bottom left: AUC performance of the different algorithms over the 50 target-pairs ordered by their corresponding AVE bias values. Right: scatter plots of the correlation between AUC and AVE bias for each of the classification algorithm over the 50 target-pairs.

We also note that inductive and confirmation bias may generate closely correlated molecules. At the beginning of a medicinal chemistry project, many evaluated molecules are inactives. The few actives are chosen for deeper exploration, and a large set of analogues are synthesized. In many discovery projects, medicinal chemistry efforts cluster near the current best molecules[9]. Therefore, when we pick an arbitrary temporal cut-off between these phases, the validation actives are often related variants of actives already discovered in the early-phase training data. Indeed, Kearnes et. al[13] showed that a significant number of highly-similar (and even identical) molecules appear in temporally-split industrial data sets. This data redundancy may artificially inflate model performance and may explain why we observe that time-split data contains significant amounts of bias. The AUCs achievable by Random Forest has an $r^2$ of 0.64 for the ChEMBL time-split.

**Overfitting**

Overfitting is a ubiquitous concern, prompting the use of held-out test sets to assess likely predictive performance. Although ML algorithms minimize training error, reducing training error to zero seldom leads to the best prospective performance[45]. For example, techniques such as regularization and early stopping are used to improve generalization despite the fact that they increase training error. Unfortunately, we cannot know *a priori* how much training error is optimal, and we can identify overfitting only by noting the divergence between training and held-out predictive performance. Furthermore, when the training and held-out datasets correlate, we may think that our models continue to improve while true prospective error rises.

The 1-NN algorithm memorizes the training set and, therefore, trivially achieves perfect performance on training data. Any gap between training and held-out data is, by definition, evidence of overfitting. Our 1-NN results show that such overfitting occurs in every benchmark we investigated. The results also show that this performance gap is not due to inherent complexity of the validation data; for example, on the J&J ECFP4 benchmark, the performance of Random Forest with an AUC of 0.85 exceeds the performance of 1-NN with an AUC of 0.76. As AVE bias predicts when 1-NN performs well, it provides a measure of the extent to which a given benchmark can be solved through overfitting. In our experience, machine learning models overfit unless explicit steps are taken to avoid it. Therefore, our opinion is that, unless we can refute the hypothesis that a model's performance is due to overfitting, it is the likeliest explanation for good results. Indeed, this is the reason practitioners report performance on held-out data in the first place.

**Unbiasing**

To avoid selecting models that are prone to overfitting, one might design a benchmark that excludes high bias cases, to focus on the low bias tests. However, excluding test cases entirely can remove large quantities of data, and data is always scarce. While our principle focus in this paper is on the construction of predictive benchmarks, a secondary concern is how to use as much available data as possible. Figure 16 illustrates that naive exclusion criteria can eliminate too much data, *e.g.* by ignoring any data point within a fixed distance from point **a**, including points on the border of different Voronoi cells. For cases where available data is scarce, we proposed an algorithm that minimizes AVE bias in test cases. The algorithm is described in the Supporting Information and uses a genetic algorithm to partition the available data into training and validation subsets with reduced bias. The algorithm is a heuristic search, and we claim neither that the algorithm is optimal nor that it is guaranteed to succeed. There is a risk that new biases are introduced into the test, as with clustering and MUV unbiasing. However, in many cases, test cases can be successfully unbiased and used in evaluations. For example, we succeeded in unbiasing approximately 400 of the 560 test cases in the J&J benchmark. As discussed in the Supporting Information, these unbiased tests appear to be more difficult than the native benchmarks for machine learning approaches. Unbiasing techniques that account for inactive-inactive bias, such as AVE bias, may provide a more accurate assessment of prospective predictive performance.

**Features**

One of the inputs to our analysis is the distance/similarity measure, so our results are parameterized by the feature definition used. If one has access to the ideal feature – that is, the label itself – then unbiasing is neither possible nor desirable. Benchmark features should be chosen to be weak enough that we don't believe *a priori* that any one feature (or small group of features) provides a signal strong enough to solve the benchmark. For example, atom counts or molecular weight should be insufficient to distinguish active molecules from inactive (and we interpret the discriminatory power of molecular weight as a benchmark design error[11]). Fortunately, similarities that are defined by the presence, absence, or count of atoms or substructures are insufficient to capture the nonadditivity of functional group contributions as observed in entropic-enthalpic tradeoffs, activity cliffs, and magic methyls[41;46]. These substructure feature sets include many commonly-used fingerprinting methods, including ECFP[22], MACCS[27], MUV simple fingerprints[17], and APFP[37;38].

**Human learning**

Although we focus on machine learning algorithms in this paper, most cheminformatics systems are human-in-the-loop learning systems and our observations also apply there. If a system admits iterative manual selection or tuning of parameters, then learning can happen implicitly. For example, forcefield-based docking algorithms do not (directly) use bioactivity data for their parameterization, but such "physics-esque" systems often have tunable parameters, such as user-defined van der Waals radii. Scoring functions where parameters are manually added to improve performance on targets of interest[11] or where pose sampling is constrained manually to regions of interest may overfit to training data as we describe here. A human operator who tunes these settings until good performance is achieved on a given test can achieve undetected overfitting without the need of machine learning algorithms.

## Conclusion & Recommendations

Low training error is insufficient justification for choosing a machine learning model, since good performance might be due to overfitting rather than accuracy. As there are many functions which overfit data and few which correctly generalize to new data, our prior belief should be that any given low-training-error model is likely to prove inaccurate. Therefore, it is common practice to pick a model only after evaluation on a held-out validation data set, and to apply clustering or MUV unbiasing to ensure that the validation set is challenging[45]. However, these previous unbiasing techniques were typically designed without consideration of machine learning methods, which can learn from either the active or inactive class and which use both a training and a validation set. We have introduced a new measurement, AVE bias, with which to evaluate the difficulty of a training-validation data split.

We have shown that tests with high AVE bias are easy to solve. AVE bias is predictive of performance across common benchmarks, standard machine learning algorithms, and a variety of molecular fingerprints. When we include inactives in the analysis, we show that AVE bias can identify the test cases which will be easy for a variety of machine learning algorithms, even when the active-active distances have been balanced as in the MUV sets. In fact, we demonstrate that the analysis is robust even when the standard techniques enumerated above have already been used to make the benchmark challenging, including clustering, approximate temporal splitting, and balancing active-active distances as in the MUV sets.

We demonstrated that test classes can be easily separable when the validation actives are more similar to training actives than to training inactives. If the relative distances among actives is smaller than between actives and inactives, machine learning algorithms can find a separating hyperplane. Symmetrically, machine learning algorithms can exploit relative similarity among inactives to solve test cases. This can lead to surprisingly easy tests, even when the molecules in the training and validation appear to be different. We demonstrated that the relative clustering of actives and inactives is highly predictive of machine learning performance. The performance correlation remains even after a variety of unbiasing and redundancy removal techniques are applied to the benchmark sets. We have also shown that removing the AVE bias from a benchmark will eliminate much of the performance of modern machine learning approaches.

These results are consistent with the interpretation that the recent apparent gains in accuracy in ligand-based screening are due to machine learning-based techniques memorizing the similarity among benchmark inactives. In turn, this conclusion would provide a unified explanation for a number of strange observations in cheminformatics. First, that benchmark performance is high while (in our experience) real-world utilization of cheminformatics is low. Second, that the performance of chemical fingerprints – even weak fingerprints like heavy atom count or molecular weight – are strongly correlated with each other. Third, that the performance of machine learning algorithms are strongly correlated with each other. Each of these observations is to be expected if we evaluate predictive systems on test sets where the test classes are easily separable.

We conclude with several recommendations. We advise, before running a performance assessment, that we first assess the strength of the proposed benchmark. Concretely, we suggest that cheminformaticians measure and report biases in their tests, and we provide our code to compute AVE biases. Tests with high bias can be excluded from consideration or unbiased with the provided code. Baseline predictive performance, such as from 1-NN, should also be included in results. We also note that structure-based technologies provide a way to model prospective testing by removing all active molecules for a target from the training data, so long as care is taken to avoid contamination from data for closely-related targets. Finally, prospective discovery testing is the most convincing test that can be performed and corresponds precisely to the intended use of such tools.

## Acknowledgments

We are very grateful to Bob Sheridan for his help with launching the initial investigation that grew into this paper, and providing the Merck data, useful discussion and suggestions. We also thank our colleagues at Atomwise, without whom this paper would not have been possible.

## Author Information

The authors declare no competing financial interest.

## Supporting Information

Supporting Information Available: Python scripts for evaluating and removing AVE bias from ligand-based benchmarks. This material is available free of charge via the Internet at `http://pubs.acs.org`.

**TOC**

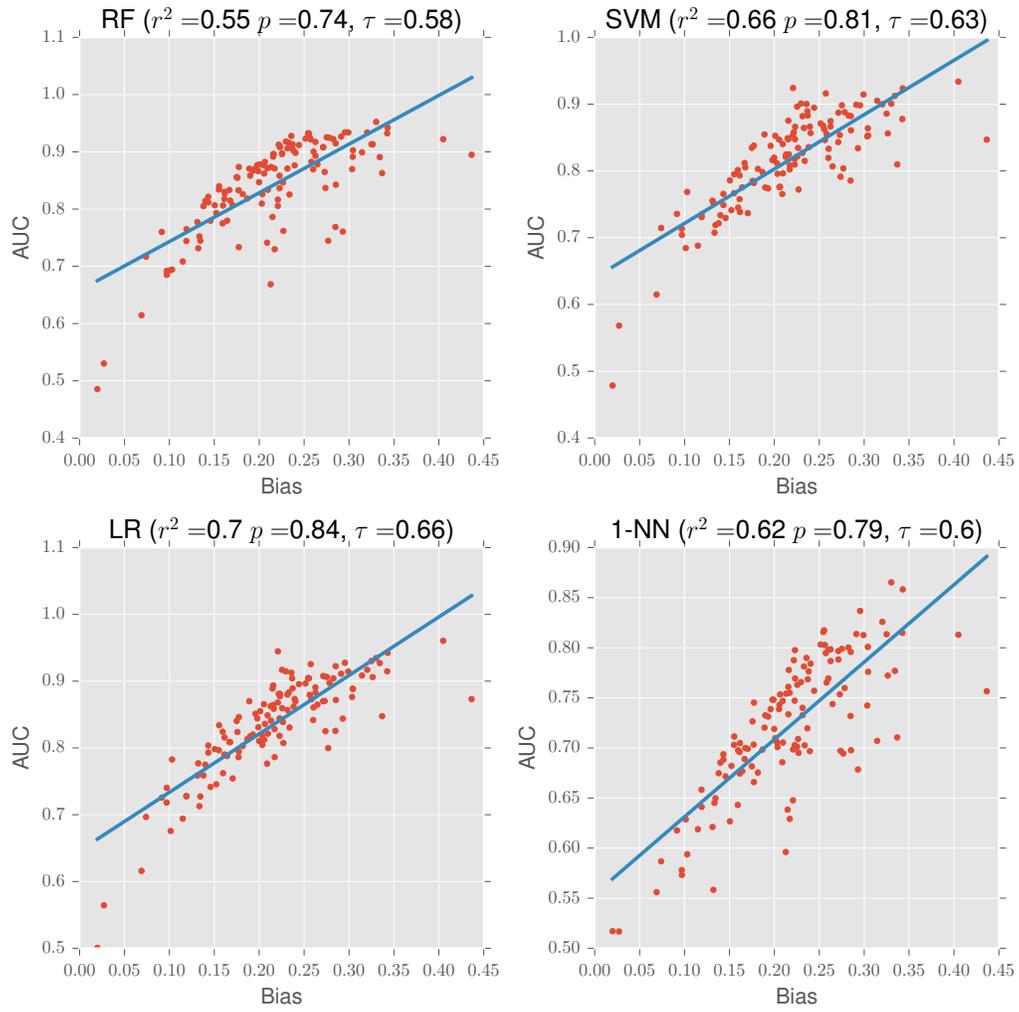

Figure 19: TOC